\documentclass[twocolumn,pre,superscriptaddress,aps,footinbib]{revtex4}
\usepackage{url,psfrag,graphicx}
\usepackage{dcolumn}
\usepackage{amsmath,amssymb}
\usepackage{bm}
\usepackage{hyperref}
\usepackage{float,epsfig,color}
\usepackage{times}
\usepackage[latin1]{inputenc}
\usepackage{graphicx}
\usepackage{amsmath}
\usepackage{amssymb}
\usepackage{dcolumn}

\begin{document}

\title{The derivative of the Kardar-Parisi-Zhang equation is not in the KPZ universality class}
\author{Enrique Rodr\'{\i}guez-Fern\'andez}
\email{enrodrig@math.uc3m.es}
\affiliation{Departamento de Matem\'aticas and Grupo Interdisciplinar de Sistemas Complejos (GISC)\\ Universidad Carlos III de Madrid, Avenida de la Universidad 30, 28911 Legan\'es, Spain}
\author{Rodolfo Cuerno}
\email{cuerno@math.uc3m.es}
\affiliation{Departamento de Matem\'aticas and Grupo Interdisciplinar de Sistemas Complejos (GISC)\\ Universidad Carlos III de Madrid, Avenida de la Universidad 30, 28911 Legan\'es, Spain}

\begin{abstract}
The Kardar-Parisi-Zhang (KPZ) equation is a paradigmatic model of nonequilibrium low-dimensional systems with spatiotemporal scale invariance, recently highlighting universal behavior in fluctuation statistics. Its space derivative, namely the noisy Burgers equation, has played a very important role in its study, predating the formulation of the KPZ equation proper, and being frequently held as an equivalent system. We show that, while differences in the scaling exponents for the two equations are indeed due to a mere space derivative, the field statistics behave in a remarkably different way: while KPZ follows the Tracy-Widom distribution, its derivative displays Gaussian behavior, hence being in a different universality class. We reach this conclusion via direct numerical simulations of the equations, supported by a dynamic renormalization group study of field statistics.
\end{abstract}

\maketitle

The Kardar-Parisi-Zhang (KPZ) equation \cite{Kardar86} describes the space-time evolution of a scalar field $h(\mathbf{r},t)$ as
\begin{eqnarray}
 & \partial_t h = \nu \nabla^2 h + (\lambda/2) (\nabla h)^2 + \eta, & \label{eq:kpz} \\
 & \langle \eta(\mathbf{r},t)\eta(\mathbf{r}',t') \rangle =2D\delta (\mathbf{r}-\mathbf{r}') \delta (t-t'), &
\label{eq:noise}
\end{eqnarray}
where $\mathbf{r}\in\mathbb{R}^d$; $\nu, D>0$, and $\lambda$ are parameters, and $\eta$ is zero-average, Gaussian white noise. This continuum model is a landmark of current Statistical Physics \cite{Taeuber14,Livi17}, being considered even on a par with the Ising model \cite{Lenz20}. Indeed, 24 years and mathematical {\em tours de force} were required for nontrivial exact solutions to be achieved in the cases of both, the Ising model \cite{Onsager44} and the KPZ equation \cite{Sasamoto10,Amir10,Calabrese11}. The former (latter) model constitutes a paramount universality class for equilibrium (non-equilibrium) critical phenomena,
defined by universal behavior of critical exponents, correlation functions \cite{Henkel99}, and amplitude ratios \cite{Henkel08,Halpin-Healy15,Takeuchi18}. Specifically, having been originally proposed to model interface growth \cite{Kardar86}, the KPZ equation displays critical behavior recently identified in very disparate contexts, including bacterial populations \cite{Hallatschek07}, turbulent liquid crystals \cite{Takeuchi11}, non-linear oscillators \cite{Beijeren12}, stochastic hydrodynamics \cite{Mendl13}, colloidal aggregation \cite{Yunker13}, thin film deposition \cite{Almeida14,Orrillo17}, reaction-diffusion systems \cite{Nesic14}, random geometry \cite{Santalla15}, superfluidity \cite{Altman15}, active matter \cite{Chen16}, or quantum entanglement \cite{Nahum17}.

While, from the point of view of exact integrability, the Ising model is most theoretically fertile in two-dimensions (2D) \cite{Henkel99}, for KPZ this happens for $d=1$. Here, the statistics of the field have been proven to be described, depending on global constraints on system size $L$ and/or initial conditions, by some member of the Tracy-Widom (TW) family of probability distribution functions (PDF) for the largest eigenvalue of random matrices \cite{Kriechebauer10,Halpin-Healy15,Takeuchi18}, demonstrating KPZ behavior as a conspicuous instance among systems with non-Gaussian fluctuations \cite{Fortin15}. Now the universality class incorporates the field statistics, the precise flavor of the TW distribution leading to universality sub-classses in the KPZ case \cite{Kriechebauer10,Halpin-Healy15,Takeuchi18}.

Historically, a major role in delineating KPZ universality has been played by the stochastic or noisy Burgers equation,
\begin{equation}
\partial_t u = \nu \partial_x^2 u + \lambda u \partial_x u + \partial_x\eta,  \label{eq:burgers}
\end{equation}
where $\eta$ is as in Eq.\ \eqref{eq:noise}. Clearly, the space derivative of Eq.\ \eqref{eq:kpz} yields Eq.\ \eqref{eq:burgers} if $u=\partial_x h$. This relation was exploited e.g.\ in \cite{Kardar86} to seminally obtain the exact scaling exponents by adapting the earlier dynamical renormalization group (DRG) analysis \cite{Forster77} of Eq.\ \eqref{eq:burgers}, as a model of a randomly stirred fluid. The noisy Burgers equation \cite{DaPrato94,Bertini97,Gubinelli13} is a paramount system on its own, e.g.\ for fluid \cite{Frisch95} and plasma \cite{Krommes02} turbulence, or for interacting particle \cite{Spohn91} and driven-diffusive systems \cite{Taeuber14}. Actually, both 1D equations, \eqref{eq:kpz} \cite{Livi17,Krug97} and \eqref{eq:burgers} \cite{Spohn91}, share an ``accidental'' fluctuation-dissipation symmetry by which the nonlinear term does not influence the corresponding stationary solution of the Fokker-Planck equation governing the field PDF, ${\cal P}$, which becomes a Gaussian, equilibrium-like distribution, determined by the linear and the noise terms \cite{Livi17,Spohn91,Krug97}. Combined with the shared symmetry under Galilean transformations, this allows to show that the two equations share the non-trivial $z=3/2$ value for the dynamic exponent describing the power-law increase of the correlation length, $\xi(t) \sim t^{1/z}$ \cite{Krug97}. The roughness exponent $\alpha$ quantifying the scaling of the field rms deviation with system size at saturation \cite{Krug97}, $w \sim L^{\alpha}$, differs as expected ($\alpha_{\rm KPZ}=\alpha_{\rm Burgers}+1=1/2$), since $h(x) = \int^x_{x_0} u(x') \, {\rm d}x'$. Thus, Eqs.\ \eqref{eq:kpz} and \eqref{eq:burgers} are frequently considered as two equivalent descriptions of a same underlying process. However, the KPZ equation shows that Gaussian behavior for the stationary ${\cal P}$ does {\em not} imply that the height statistics prior to saturation (for $L<\infty$) are also Gaussian; indeed, they are TW-distributed for KPZ \cite{Halpin-Healy15,Kriechebauer10,Takeuchi18}.

From the point of view of the specific physical systems described by the noisy Burgers equation \cite{Taeuber14,Frisch95,Krommes02,Spohn91}, it is crucial to clarify whether their field statistics are also non-Gaussian in the growth regime, in order to accurately identify the universality class of their kinetic roughening behavior. In this Letter we show that this is not the case, i.e., we show that the one-point PDF for $u(x,t)$ as described by Eq.\ \eqref{eq:burgers} is Gaussian for times dominated by the nonlinearity, crucially prior-to and (as expected) after saturation to steady state. We reach this conclusion by direct numerical simulations of the equation, which are analytically supported by a DRG analysis of the field statistics for Eq.\ \eqref{eq:burgers}. We also address the dynamics of the space-integral of Eq.\ \eqref{eq:burgers}, explicitly illustrating that, in this case, the KPZ sum, $h(x,t)$, of (correlated) Gaussian Burgers variables $u(x,t)$ indeed yields TW statistics \cite{Clusel08}.

\begin{figure}[!t]
\begin{center}
\includegraphics[width=1.0\columnwidth]{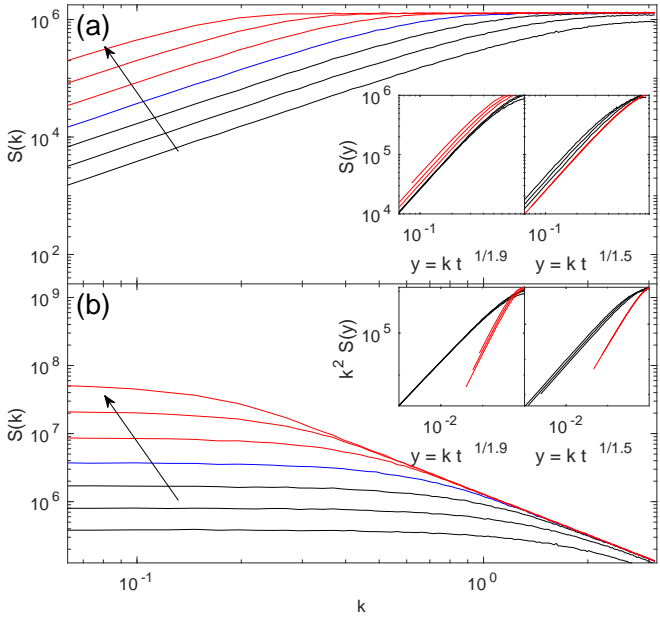}
\caption{Time evolution of the structure factor described by Eq.\ \eqref{eq:burgers} for (a) $u(x,t)$ and (b) $h(x,t)=\int_{0}^x u(x',t) \, {\rm d}x'$, using $D=\nu=1$, $\lambda=4$, and $L=256$. Black (red) solid lines correspond to the linear (nonlinear) regime, as implied by the data collapses in the insets. The arrows indicate time increase, $t$ for each line being twice that of the previous one, starting at $t_0=0.64$. All units are arbitrary.}
\label{fig:Sk}
\end{center}
\end{figure}

We begin by addressing the full time dynamics described by Eq.\ \eqref{eq:burgers}. While the invariant measure of the equation has been shown \cite{Spohn91,DaPrato94,Bertini97} to be Gaussian, and the asymptotic scaling exponents are analytically known via DRG \cite{Forster77,Ueno05}, to our knowledge the time crossover which occurs from linear to nonlinear behavior has not been explicitly addressed yet. In order to assess it, we have performed numerical simulations of Eq.\ \eqref{eq:burgers}. Note, this model is known to be conspicuously prone to numerical instabilities \cite{Hairer11}. We use the numerical scheme proposed in \cite{Sasamoto09}, which provides consistent results. Considering flat initial conditions and periodic boundary conditions, we show in Fig.\ \ref{fig:Sk} the time evolution of the structure factor
$S(k,t)=\langle \tilde{\phi}(k,t) \tilde{\phi}(-k,t) \rangle $, as described by Eq.\ \eqref{eq:burgers}; here, tilde denotes spatial Fourier transform and $k$ is wave number. Panel (a) corresponds to $\phi(x,t)=u(x,t)$, while panel (b) is for its space integral, $\phi(x,t)=h(x,t)=\int_0^x u(x',t) \, {\rm d}x'$, which should retrieve the behavior expected for Eq.\ \eqref{eq:kpz}. At relatively early times, the linear term and the noise in Eq.\ \eqref{eq:burgers} are expected to control the evolution of both the $u$ and $h$ fields, hence $z=2$ as provided by the exact solution of the linearized equation \cite{Krug97}. This behavior is approximately reproduced by our simulations, as implied by the data collapse shown in the insets for small times. Indeed, recall that, under kinetic roughening conditions, $S(k,t) \sim k^{-(2\alpha+1)} s(kt^{1/z})$, with $s(u)\sim 1$ for $u\gg 1$ and $s(u)\sim u^{2\alpha+1}$ for $u\ll 1$ \cite{Barabasi95,Krug97}. Collapse is achieved for $u$($h$) using $\alpha=-1/2$ (1/2), as also borne out from the exact solution of the linearized equations \eqref{eq:burgers} and \eqref{eq:kpz}, respectively. However, for sufficiently long times, the value of $z$ changes, indicating nonlinear behavior. Indeed, data collapse is now obtained using $z=3/2$ both, for $u$ and for $h$, as expected in the asymptotic limit \cite{Forster77,Kardar86}. Note that, also in both cases, $\alpha$ remains fixed to its linear-regime value as a consequence of the ``accidental'' fluctuation-dissipation symmetry \cite{Spohn91,Krug97,Livi17}. Overall, Eq.\ \eqref{eq:burgers} is thus seen to account for the full dynamics of the Burgers field, and for the KPZ behavior of its space integral. Conversely, in the Supplemental Material (SM) (see Appendix \ref{SM}) we integrate numerically the KPZ equation \eqref{eq:kpz} showing that the evolution of its slope field $u(x,t) = \partial_x h(x,t)$ coincides with results from Eq.\ \eqref{eq:burgers}. All this supports the consistency of our numerical results.

%\begin{figure*}[!t]
\begin{figure}[!t]
\begin{center}
%\hspace*{-0.8cm}
\includegraphics[width=1.0\columnwidth]{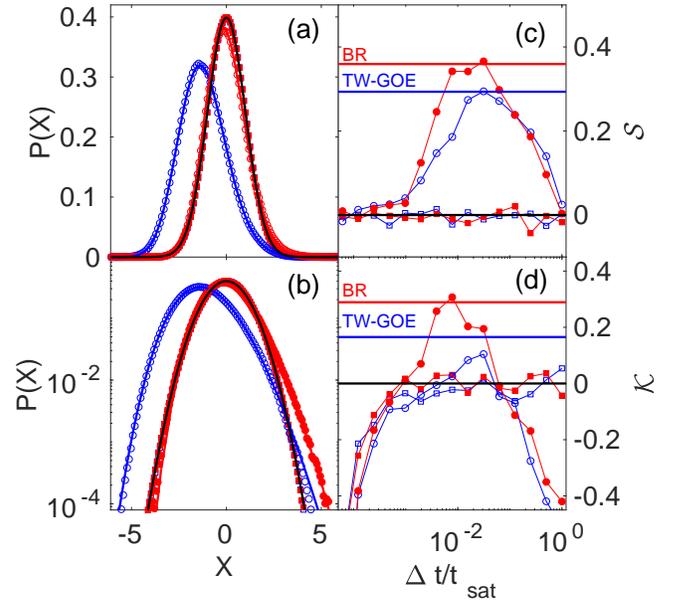}
\caption{Histograms for $X (x,\Delta t,t_0)$ (a,b) from  simulations of Eq.\ \eqref{eq:burgers} with $\nu=1$, $\lambda=10^3$, $D=10^{-3}$, and $L=256$, for $\phi=u$ (squares) and $\phi=h(x,t)=\int_{0}^x u(x',t) \, {\rm d}x'$ (circles). Means and variances have been adjusted to TW-GOE and BR values. Dynamics of skewness (kurtosis) appears in (c) [(d)]. In all panels blue (red) and empty (filled) symbols correspond to the growth (saturation) regime, with black, blue, and red solid lines showing exact Gaussian, TW-GOE, and BR values, respectively; $t_{sat}=100$, and $\Delta t=25-50, 1.5-3$, and $0.4-0.8$ are used for Gaussian, TW-GOE, and BR-like histograms, respectively. Thin lines in (c,d) are guides to the eye. All units are arbitrary.}
\label{fig:Histograma}
\end{center}
\end{figure}
%\end{figure*}

Beyond scaling exponents, we have also studied numerically the field statistics described by Eq.\ \eqref{eq:burgers}, by considering
\begin{equation}\label{eq:X}
    X (x,\Delta t,t_0) = (\Delta \phi - \overline{\Delta \phi})/(\Gamma \Delta t)^{\beta},
\end{equation}
where $\Delta \phi (x,\Delta t,t_0)= \phi (x,t_0+\Delta t) - \phi (x,t_0)$, bar denotes space average, $\beta=\alpha/z$ is the growth scaling exponent, $\Gamma$ is a normalization constant \cite{Takeuchi13}, and $\Delta t \gg 1$ will be assumed. In principle, the statistical distribution of $X (x,\Delta t,t_0)$ can differ before ($t_0=0, \Delta t \ll t_{sat}$) and after ($t_0 > t_{sat}$) saturation. E.g., for a periodic KPZ system, they are provided by the TW distribution for the largest eigenvalue of random matrices in the Gaussian Orthogonal Ensemble (TW-GOE) and by the Baik-Reins (BR) distributions, respectively \cite{Kriechebauer10,Takeuchi13,Halpin-Healy15,Takeuchi18}.

We assess in Fig.\ \ref{fig:Histograma} the histogram of $X (x,\Delta t,t_0)$ for the $u$ and $h$ fields, as numerically obtained from Eq.\ \eqref{eq:burgers}. Full PDFs are shown in Figs.\ \ref{fig:Histograma}(a) and \ref{fig:Histograma}(b) for times both in the nonlinear growth regime determined above ($t_0+\Delta t<t_{sat}$) and after saturation to steady state ($t_0>t_{sat}$). Figures \ref{fig:Histograma}(c) and \ref{fig:Histograma}(d) show the time evolution of the field skewness, $\mathcal{S}=\langle X^3 \rangle_c/\langle X^2 \rangle_c^{3/2}$ and excess kurtosis, $\mathcal{K}=\langle X^4 \rangle_c/\langle X^2 \rangle_c^{2}$, respectively, where $\langle X^n \rangle_c$ denotes the $n$-th order cumulant. The statistics of $u(x,t)$ are Gaussian to a high precision, both prior to and after saturation, see the PDFs in panels (a,b). Indeed, the skewness and (somewhat more slowly) the excess kurtosis converge rapidly to zero [panels (c,d)] for $u(x,t)$. The slope field of Eq.\ \eqref{eq:kpz} exhibits a similar Gaussian behavior, as shown in Fig.\ \ref{fig:Histograma} of the SM (Appendix \ref{SM}), again supporting the identification of solutions of Eq.\ \eqref{eq:burgers} with the slopes field for Eq.\ \eqref{eq:kpz}.

In the case of the $h(x,t)$ field, Eq.\ \eqref{eq:burgers} correctly leads $\mathcal{S}(t)$ and $\mathcal{K}(t)$ to take on the characteristic universal values of the KPZ equation, either TW-GOE or BR [shown as blue or red solid lines, respectively, in Figs.\ \ref{fig:Histograma}(c) and (d)] for intermediate values of $\Delta t$ within the expected ranges of $t_0$ and $\Delta t$ ($t_0=0$, $t_{sat} > \Delta t \gg 0 $ and $t_0>t_{sat}$, $t_0 > \Delta t \gg 0$, respectively). Indeed, the PDF of $h$ fluctuations approaches the TW-GOE or BR distributions for $t_0=0$ or $t_0=t_{sat}$, respectively, only for such intermediate values of $\Delta t$. This behavior has been also observed for discrete and continuum models in the KPZ universality class \cite{Takeuchi13,Squizzato18}. Specifically, the difference between the actual PDF and the ideal TW-GOE or BR distributions reaches a minimum for intermediate values of $\Delta t$. It is for such $\Delta t$ that the numerical $h$-PDF is plotted in Figs.\ \ref{fig:Histograma}(a,b). Means and variances have been adjusted to equal those of the exact TW-GOE or BR distributions. As the pre- or post-saturation $h$-PDF evolves from Gaussian to TW-GOE or BR, to become Gaussian again for large $\Delta t$ [see panels \ref{fig:Histograma}(c,d)], the Gaussian black solid line on panels \ref{fig:Histograma}(a,b) seems to attract the tails of the $h$ distribution.

\begin{figure}[!t]
\begin{center}
\includegraphics[width=1\columnwidth]{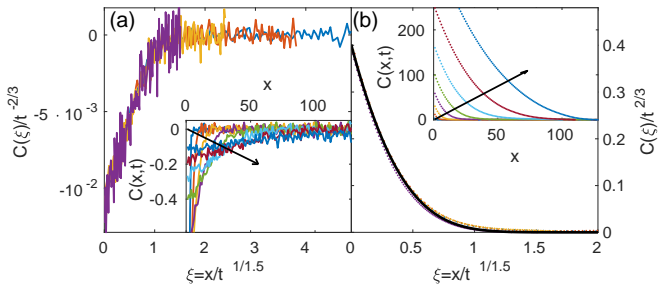}
\caption{Collapse of the two-point correlation function $C(x,t)$ at different times described by Eq.\ \eqref{eq:burgers} for (a) $\phi=u$ and (b) $\phi=h=\int_{0}^x u(x',t) \, {\rm d}x'$, for $L=256$, $\Delta t = 10^{-2}$, $\nu=D=1$, $\lambda=10$. Insets show the uncollapsed data. The small range of values for $C$ induces large relative errors in (a). The solid line in (b) shows the exact covariance of the Airy$_1$ process \cite{Bornemann10}. Arrows show time increase, with $t$ doubling for each line, from $t_0=1$. All units are arbitrary.}
\label{fig:Cbefore}
\end{center}
\end{figure}

A final stark difference in the critical behavior of the $u$ (Gaussian) and $h$ (KPZ) fields, as described by Eq.\ \eqref{eq:burgers}, lies in the behavior of the two-point correlation function $C(x,t) = \langle \phi(x_0,t)\phi(x_0+x,t)\rangle-\langle \bar{\phi}(t) \rangle^2$, see Fig.\ \ref{fig:Cbefore}. Although in both cases the expected scaling form holds \cite{Barabasi95,Krug97}, $C(x,t) = t^{2\beta} c(x/t^{1/z})$ with $c(y)\sim {\rm cst.}-y^{2\alpha}$ for $y\ll 1$ and $0$ for $y\gg 1$, the exponents leading to collapse are different [i.e., those derived from Fig.\ \ref{fig:Sk}], as are the corresponding scaling functions $c(y)$. For the $h$ field, the latter is the covariance of the Airy$_1$ process, as expected in the growth regime for 1D KPZ scaling with periodic boundary conditions \cite{Bornemann10,Halpin-Healy15,Takeuchi18}. {\em Qualitative} differences between Burgers and KPZ behaviors seem larger for the one-point statistics than for $C(x,t)$.

\begin{figure}[!t]
\begin{center}
\includegraphics[width=0.8\columnwidth]{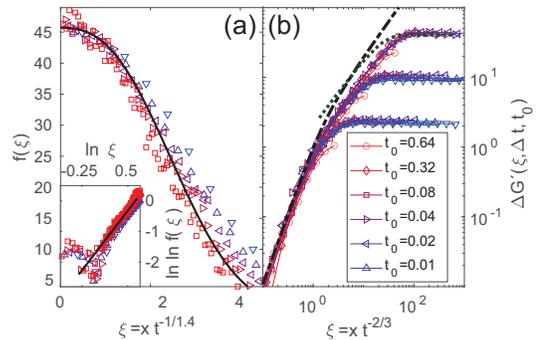}
\caption{Two-point correlation function $G(x,t)$ from numerical simulations of Eq.\ \eqref{eq:burgers} for (a) $\phi=u$ and (b) $\phi=h=\int_{0}^x u(x',t) \, {\rm d}x'$, for $L=2^{10}$, $\Delta t=5 \cdot 10^{-3}$, $\nu=D=1$, $\lambda=6$. In panel (a) the black solid line is a fit to $f(\xi) \propto e^{-a_2 \xi^{2.6}}$, and has slope 2.6 in the inset. The exponents required for collapse are close (but not identical) to the theoretical expectations \cite{Takeuchi18,Prahofer04,Imamura12}. In panel (b) $\Delta G'=G'(\xi,\Delta t, t_0)-G'(0,\Delta t, t_0)$, the black dash-dotted line is the stationary KPZ correlation $g(\xi)-g(0)$, and the green dotted line is the Airy$_1$ correlation, as in \cite{Takeuchi13}. In both panels, color evolves from blue to red for increasing $t$ (a) or $t_0$ (b). The number of realizations is larger in (a) ($2 \cdot 10^4$) than in (b) ($10^2$). All units are arbitrary.}
\label{fig:Cafter}
\end{center}
\end{figure}

Our simulations of Eq.\ \eqref{eq:burgers} likewise reproduce the expected two-point correlations for $u$ and $h$ {\em after saturation} to steady state, as shown by studying $G(l,\Delta t,t_0) = \langle (\delta\phi(x+l,t_0+\Delta t)-\delta\phi(x,t_0))^2 \rangle$ with $\delta\phi=\phi-\langle \phi\rangle$. For $\phi=u$, one expects \cite{Prahofer04,Imamura12,Takeuchi18} $G(x,t) \propto t^{1/z} f(a_1 x/t^{1/z})$, for $t_0>t_{sat}$, where $f(y) \sim e^{-a_2 y^3}$ with $a_{1,2}$ being numerical constants. In turn, for $\phi=h$ and large $t_0$ and $\Delta t$, one expects $G'(\xi,\Delta t,t_0) \equiv t^{-1/z} G(\xi,\Delta t,t_0) \simeq g(\xi)$, with $\xi=l/t^{1/z}$ and $g(\xi)$ the exact solution for the rescaled stationary KPZ correlation, which is such that $f(\xi)\propto g''(\xi)$ \cite{Prahofer04,Imamura12,Takeuchi13}. This is indeed the behavior found in Fig.\ \ref{fig:Cafter}(b), quite similar to that assessed in \cite{Takeuchi13} for a discrete model in the 1D KPZ class. Note again that fluctuations are much stronger for $\phi=u$ than for $\phi=h$.

The Gaussian behavior obtained for Eq.\ \eqref{eq:burgers} coincides with analytical DRG results which follow successful analyses of field statistics for the KPZ \cite{Singha14,Singha15,Singha16} and nonlinear-Molecular Beam Epitaxy (MBE) \cite{Singha16b} equations, and for the scalar Burgers equation with \emph{non}-conserved noise \cite{Rodriguez19}. The method performs a partial RG transformation in which the equation is coarse-grained \cite{Yakhot86}, while omitting the rescaling step \cite{Forster77,Ueno05}. Within a one-loop approximation \cite{Singha14,Singha15,Singha16,Singha16b,Rodriguez19} [see details in the SM (Appendix \ref{SM})],
$
w^2 = \langle u^2 \rangle_c \propto \int_{\mathbb{R}} dk_1 .
$
The variance of $u$ scales as $1/L$ for $L\gg s$, with $s$ the lattice spacing,
which agrees with the expected $\alpha=-1/2$. Moreover, $w^2 \sim s^{-1}$ for $s \ll 1$ \cite{Krug97,Smith17}. We further obtain (see SM in Appendix \ref{SM}) $\langle u^4 \rangle_c \sim [\ln(1/s)]^{0.79}$, hence the kurtosis, $\mathcal{K} \to 0$ as $s \to 0$. Finally, as in \cite{Rodriguez19}, an exact symmetry induces $\langle u^{2n+1} \rangle_c=0$ $\forall n$, hence $\mathcal{S} = 0$. More generally, the $u$-PDF is symmetric (unlike the TW or BR distributions) \cite{Rodriguez19}. Combined with the vanishing kurtosis, these results fully agree with the Gaussian statistics numerically found for $u$.

Our numerical and analytical results indicate that the long-time behavior of Burgers equation with conserved noise, Eq.\ \eqref{eq:burgers}, albeit controlled by the nonlinear term, displays Gaussian statistics. This is in spite of the fact that it is precisely such nonlinearity which breaks the inversion symmetry ($u \leftrightarrow -u$) of the equation. This lack of symmetry has been correlated in the KPZ \cite{Krug97} and nonlinear-MBE equations \cite{Carrasco16} with a non-zero skewness due to the existence of a preferred growth direction \cite{Barabasi95}. Hence, the symmetry of the (Gaussian) PDF is an emergent property of the large-scale behavior in Eq.\ \eqref{eq:burgers}, much as it is for Burgers equation with {\em non-conserved} noise (NCN) \cite{Rodriguez19}. Akin to the latter, the symmetric field PDF in the nonlinear regime can be related with the behavior of the deterministic (viscous) Burgers equation, which is analytically known \cite{Burgers74,Bendaas18} to yield sawtooth profiles, symmetric, as Eq.\ \eqref{eq:burgers}, under a combined $(x,u)\leftrightarrow (-x,-u)$ transformation. This nonlinear behavior can be specifically assessed in the slopes histogram, again as in Burgers equation with NCN \cite{Rodriguez19,Sakaguchi03}, being enhanced for large $\lambda$ and small $\nu$ and $D$ values. Figure \ref{fig:Histograma_3} of the SM (Appendix \ref{SM}) show the time evolution of the $u$ slopes ($\partial_x u$) statistics, which evolves from a symmetric PDF in the early linear regime to a non-symmetric form in the nonlinear regime, to finally symmetric again at saturation. Once again, similar behavior is observed in numerical simulations of the KPZ equation, Eq.\ \eqref{eq:kpz}, now for the \emph{curvatures} ($\partial_x^2h$) of the KPZ profiles, see Fig.\ \ref{fig:Histograma_2} in the SM (Appendix \ref{SM}).

We also prove in the SM (Appendix \ref{SM}) that Eq.\ \eqref{eq:burgers} is analogous to Burgers equation with NCN \cite{Rodriguez19} with respect to two additional issues, namely: {\em (i)} It admits a scalar 2D extension, for which we obtain a symmetric PDF and the same scaling exponents as in 1D, suggesting that $d=2$ is above the critical dimension in this universality class; {\em (ii)} the nonlinear Eq.\ \eqref{eq:burgers} admits an {\em exact} linear approximation, i.e., a linear equation can be formulated with the same scaling exponent values and field PDF as Eq.\ \eqref{eq:burgers}.

In summary, we have obtained that the field statistics of Eq.\ \eqref{eq:burgers} are Gaussian, in spite of the facts that its asymptotic behavior is controlled by a nonlinear term which explicitly breaks the up-down symmetry and that the equation is related to KPZ through a mere space derivative. Such non-symmetric statistics indeed occurs both, for the integral and the slope fields related with the $u$ field obeying Eq.\ \eqref{eq:burgers}. In particular, all this behavior provides a nontrivial example in which the KPZ sum, $h(x,t)=\int_{x_0}^{x} u(x',t) \, {\rm d}x'$, of (correlated) Gaussian Burgers variables $u$ yields non-Gaussian KPZ variables $h$; albeit counterintuitive, this effect is not unknown \cite{Stoyanov97}. The correct identification of the universality class (including scaling exponent values and field PDF) is paramount to fully identify stochastic Burgers behavior in the many contexts of spatially-extended systems, from fluid turbulence to driven diffusive systems, in which Eq.\ \eqref{eq:burgers} plays a relevant role as a physical model.

\begin{acknowledgments}
We acknowledge valuable comments by B.\ G.\ Barreales, M.\ Castro, J.\ Krug, P.\ Rodr\'{\i}guez-L\'opez, and J.\ J.\ Ruiz-Lorenzo. This work has been supported by Ministerio de Econom\'{\i}a y Competitividad, Agencia Estatal de Investigaci\'on, and Fondo Europeo de Desarrollo Regional (Spain and European Union) through grants Nos.\ FIS2015-66020-C2-1-P and PGC2018-094763-B-I00. E.\ R.-F.\ also acknowledges financial support by Ministerio de Educaci\'on, Cultura y Deporte (Spain) through Formaci\'on del Profesorado Universitario scolarship No.\ FPU16/06304.
\end{acknowledgments}

\newpage

\appendix

\begin{widetext}

\section{\normalsize{Supplemental material for} \\ \mbox{} \\ \large{\bf ``The derivative of the KPZ equation is not in the KPZ universality class''} \\ \mbox{} \\ \normalsize{Enrique Rodr\'{\i}guez-Fern\'andez and Rodolfo Cuerno}}\label{SM}

\subsection{Noisy Burgers equation as the derivative of the KPZ equation}

To further assess the relation between Burgers equation with conserved noise,
\begin{eqnarray}\label{Burgers}
 & \partial_t u = \nu \partial_x^2 u + \lambda u \partial_x u + \partial_x\eta, &
    \\
 & \langle \eta(x,t)\eta(x',t') \rangle =2D\delta (x-x') \delta (t-t') , &  \nonumber
\end{eqnarray}
and the Kardar-Parisi-Zhang (KPZ) equation,
\begin{eqnarray}\label{KPZ}
 & \partial_t h = \nu \partial_x^2 h + (\lambda/2) (\partial_x h)^2 + \eta, &
    \\
 & \langle \eta(x,t)\eta(x',t') \rangle =2D\delta (x-x') \delta (t-t'), & \nonumber
\end{eqnarray}
namely, that the former is the space derivative of the latter, here we simulate numerically both, Eq.\ \eqref{Burgers} and \eqref{KPZ}, taking the space derivative of the latter for each time and noise realization, compute the structure factor for both numerical fields, and compare the results. Recall that the stochastic nonlinear equations which we are discussing are conspicuously prone to numerical inaccuracies and instabilities \cite{Sasamoto09,Hairer11}, which renders nontrivial the present type of check which we are performing. Results are provided in Fig.\ \ref{fig:Sk}, in which panel (a) corresponds to Eq.\ \eqref{Burgers} [thus repeating the same data shown in Fig.\ 1(a) of the main text (MT) for the reader's convenience] and panel (b) corresponds to the numerical derivative of the KPZ profile described by Eq.\ \eqref{KPZ}. As expected, results are virtually indistinguishable, hence consistent with the behavior discussed in the MT for the Burgers equation with conserved noise, Eq.\ \eqref{Burgers}, namely, early-time (linear regime) exponent values $z_{\rm linear}=1.9$, $\alpha_{\rm linear}=-1/2$ and late-time (nonlinear regime) exponent values $z_{\rm nonlinear}=3/2$, $\alpha_{\rm nonlinear}=-1/2$.

We proceed similarly to compute the probability distribution function (PDF) of the field 
[using Eq.\ (4) of the MT] both, for Eq.\ \eqref{Burgers} and for the numerical derivative of Eq.\ \eqref{KPZ}. Results are provided in Fig.\ \ref{fig:Histograma}. The histograms have been computed for the same parameter conditions as in Fig.\ 2 of the MT, both for the same $t_0$ and $\Delta t$ values for which Tracy-Widom (TW) and Baik-Rains (BR) distributions are obtained for the $h$ field described by Eq.\ (3) there. The histograms shown in Figs.\ \ref{fig:Histograma}(a),(b) are Gaussian to a high precision, compare the symbols in the figures with the exact Gaussian forms (solid lines). Also, the skewness and kurtosis shown in Figs.\ \ref{fig:Histograma}(c) and \ref{fig:Histograma}(d), respectively, are seen to readily take on their Gaussian (zero) values. All these results support the interpretation of the noisy Burgers equation as the derivative of the KPZ equation, as well as the Gaussian behavior of its fluctuations both, prior and after saturation to steady state, as assessed by our numerical simulations.

\begin{figure}
\begin{center}
\includegraphics[width=1.0\columnwidth]{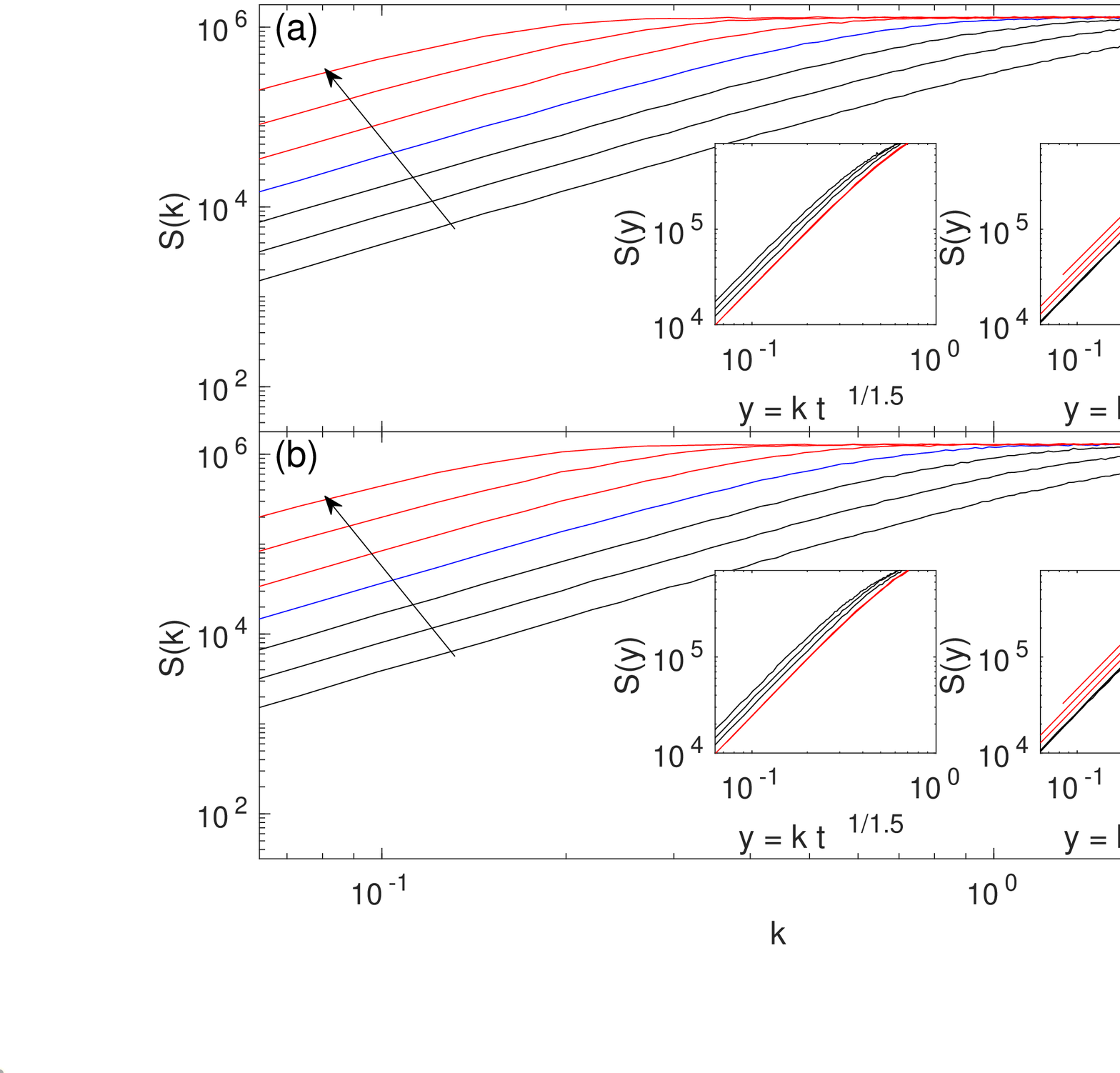}
\caption{Time evolution of the structure factor of solutions of the Burgers equation with conserved noise (a) and the slopes of the KPZ equation (b), for $D=1$, $\lambda=4$, $\nu=1$, and $L=256$. Black (red) solid lines correspond to the linear (nonlinear) regime, as implied by the collapse shown in the insets. Time increases following the arrow, $t$ for each line being twice that of the previous one, starting at $t_0=0.64$. All units are arbitrary.}
\label{fig:Sk}
\end{center}
\end{figure}

\begin{figure}[!b]
\begin{center}
\includegraphics[width=1.0\columnwidth]{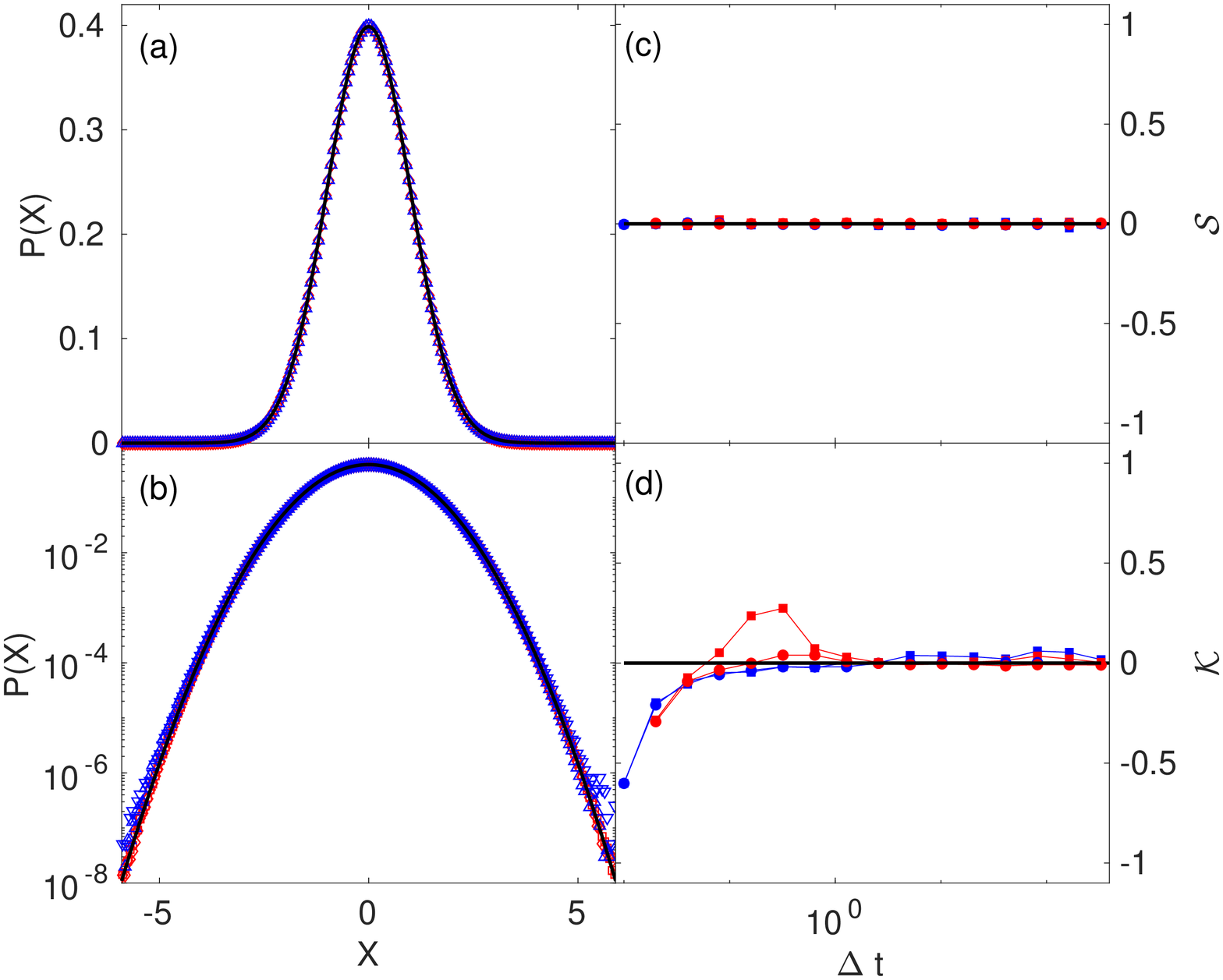}
\caption{Fluctuation histogram [using $X$ as defined in Eq.\ (4) of the MT] for $t_0=0$ (blue) and $t_0=t_{sat}=300$ (red) and $\Delta t = 150$, from numerical simulations of Burgers equation with conserved noise (squares) and from the derivative (slope field) of numerical simulations of the KPZ equation (circles), using parameters as in Figs.\ \ref{fig:Sk}(a),(b). The solid lines correspond to a Gaussian distribution. Time evolution of the fluctuations skewness (c) and kurtosis (d) for the same numerical simulations as in panels (a) and (b). All units are arbitrary.}
\label{fig:Histograma}
\end{center}
\end{figure}

\subsubsection{Asymmetric profiles}
In order to illustrate the discussion made in the MT, and in analogy with the simulations provided in Figs.\ \ref{fig:Sk} and \ref{fig:Histograma}, we assess the relevance of sawtooth-like features in the long-time behavior of the noisy Burgers equation, by evaluating the fluctuation histogram for {\em (i)} the slopes of the $u$ field from the solutions of the Burgers equation with conserved noise and {\em (ii)} the second-order space derivative (curvature field) of the $h$ field from the solutions of the KPZ equation. As expected, in both cases the profiles are asymmetric for intermediate times within the nonlinear regime [Figs.\ \ref{fig:Histograma_2} and \ref{fig:Histograma_3} (b)-(d)], away both from the linear [Figs.\ \ref{fig:Histograma_2} and \ref{fig:Histograma_3} (a)] and from the saturation [Figs.\ \ref{fig:Histograma_2} and \ref{fig:Histograma_3} (e)] regimes, in which the surface is $x\leftrightarrow -x$. symmetric on average.

\begin{figure}%[!t]
\begin{center}
\includegraphics[width=1.0\columnwidth]{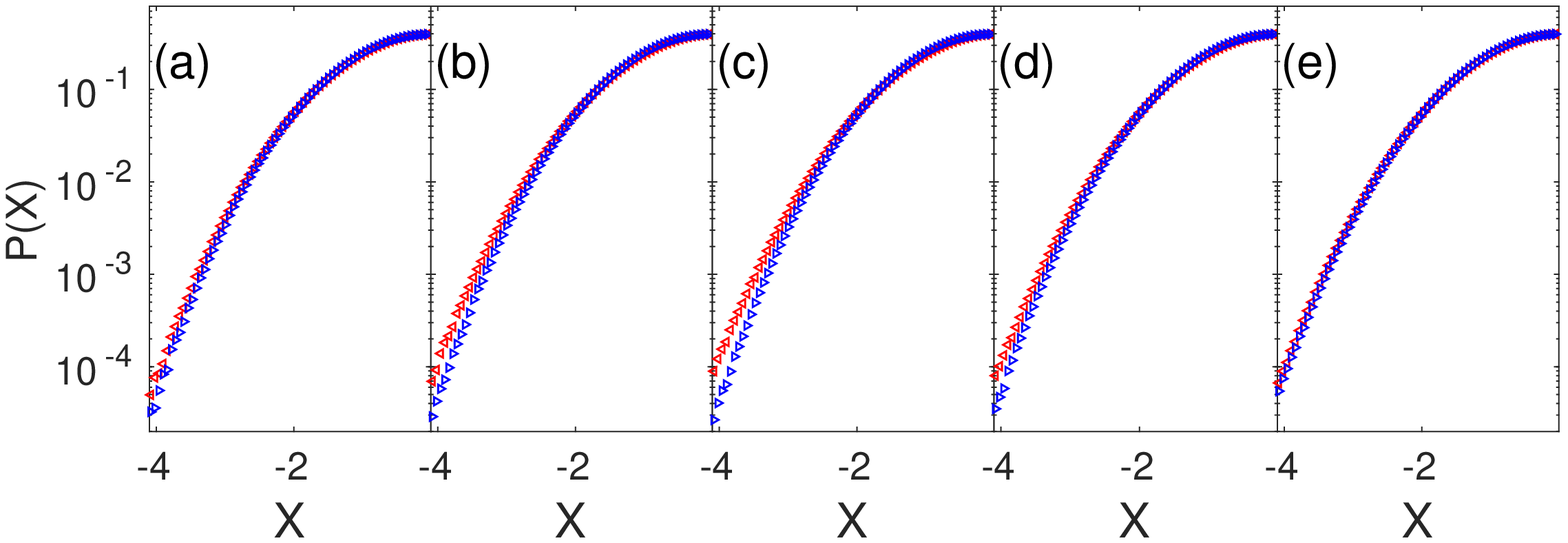}
\caption{Histogram for the slope field $\phi=\partial_x u$ [using $X=(\phi-
\bar{\phi})/{\rm std}(\phi)$] from numerical simulations of Eq.\ (3) from the MT for $\nu=1$, $\lambda=10^4$, $D=10^{-3}$, and $L=256$, for times in the linear (a), nonlinear (b), (c), (d), and saturation (e) regimes [time for each panel is twice that of the previous one, starting at $t_0 = 40$ (a)]. The $X>0$ data (red left triangles) have been reflected to facilitate comparison with $X<0$ data (blue right triangles). All units are arbitrary.}
\label{fig:Histograma_3}
\end{center}
\end{figure}

\begin{figure}
\begin{center}
\includegraphics[width=1.0\columnwidth]{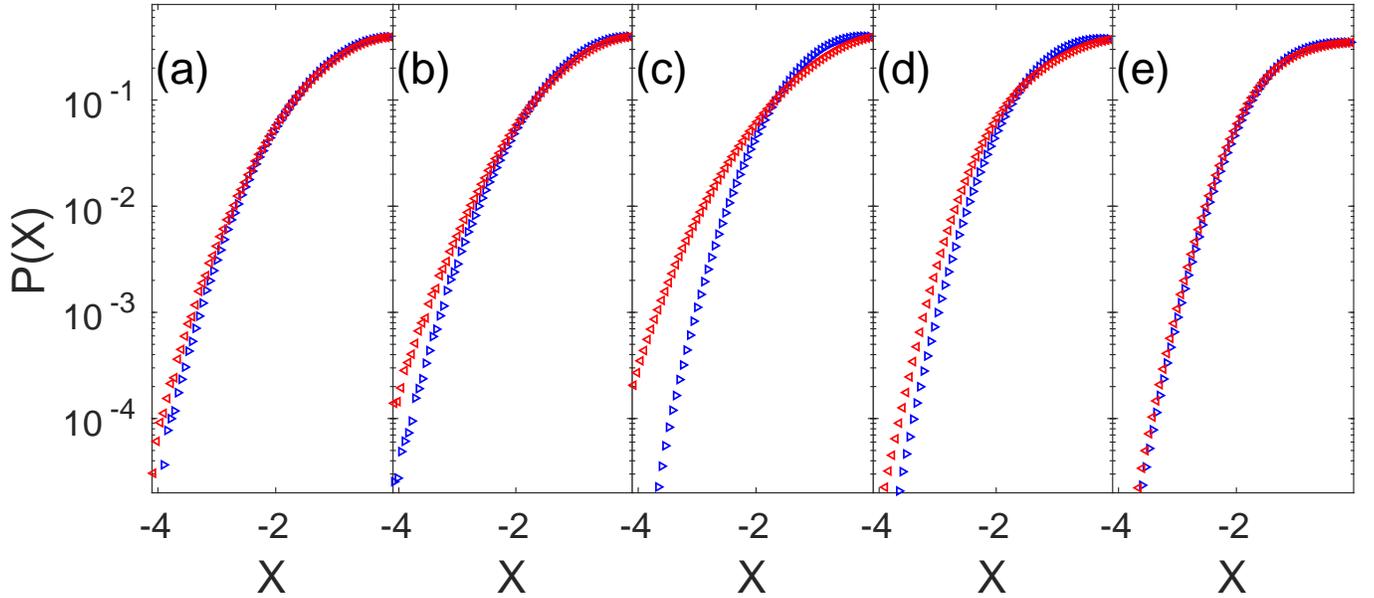}
\caption{Fluctuation histogram from numerical simulations of the curvature field (second-order space derivative) from numerical simulations of the KPZ equation, for times in the linear (a), nonlinear (b-d), and saturation (e) regimes (time for each panel is twice that of the previous one, starting at $t_0 = 40$). The histogram for $X>0$ (red left triangles) has been reflected to facilitate comparison with the $X<0$ (blue right triangles) data. Parameters as in Fig.\ \ref{fig:Sk}, except for $\lambda=10^4$ and $D=10^{-3}$. All units are arbitrary.}
\label{fig:Histograma_2}
\end{center}
\end{figure}

%%%%%%%%%%%%%%%%%%%%%%%%%%%%%%%%%%%%%%%%%%%%%%%%

\clearpage
\subsection{Dynamical Renormalization Group analysis of field statistics for the 1D noisy Burgers equation}

This section provides additional details on the evaluation of field cumulants for the noisy Burgers equation (1), following the Dynamical Renormalization Group (DRG) approach of \cite{Yakhot86}, previously applied to the evaluation of field statistics in the cases of the KPZ \cite{Singha14,Singha15,Singha16b} and the non-linear Molecular Beam Epitaxy \cite{Singha16} equations, and of the Burgers equation with non-conserved noise \cite{Rodriguez19}.

According to Eq.\ \eqref{Burgers}, the $n$-th cumulant of $u$ reads
\begin{equation}
\langle u^n \rangle_c = \int_{\mathbb{R}^{2(n-1)}} G(k_n,\omega_n) L_n  \prod_{j=1}^{n-1} \frac{dk_j d\omega_j}{(2\pi)^2} G(k_j,\omega_j),
\end{equation}
with $G(k,\omega)=(-{\rm i}\omega + \tilde{\nu}(k) k^2)^{-1}$, $\tilde{\nu}(k)= \sqrt{ \frac{\lambda^2 D}{2\pi \nu } } k^{-1/2} $ (see \cite{Singha14} for details), $G(k,\omega) \hat{\eta} = \hat{u}(k,\omega)$, hat is space-time Fourier transform, $k$ is wave-number, $\omega$ is time frequency, $k_n=-\sum_{j=1}^{n-1}k_j$, and $\omega_n=-\sum_{j=1}^{n-1}\omega_j$. The correction $L_n$ is perturbatively computed to one loop order as
\begin{equation}{}
    L_n = (2D) \delta_{n,2} + L_{n,1},
\end{equation}
where $
    L_{n,1} = K \lambda^n {\rm i}^n k_n l_{n,1} \prod_{j=1}^{n-1} k_j  $
\ is the lowest-order correction in the Feynman expansion of the cumulants, with $K=(2n-2)!!$ being a combinatorial factor (number of different fully-connected diagrams).
As we are interested in the $(k_i,\omega_i) \rightarrow (0,0)$ limit,
\begin{equation}{}
    l_{n,1}= \int_{-\infty}^{\infty} \frac{d\Omega}{2\pi} \int^> \frac{dq}{2\pi} |G_0(q,\Omega)|^{2n} (2D q^2)^n,
\end{equation}
where $G_0(k,\omega)=(-{\rm i}\omega + \nu k^2)^{-1}$ and the integration domain in $\int^>$ is the region $\{ q\in \mathbb{R} | \Lambda(\ell)=\Lambda_0 e^{-\ell} <|q|<\Lambda_0 \}$. After integration, 
%AQUI
\begin{equation}{}
    l_{n,1} = \frac{2^{n+1} \Gamma(n-\frac{1}{2})}{4 \pi^{3/2} (n-1)!} \frac{D^n \nu^{1-2n}}{\Lambda^{2n-3}(\ell)} \frac{e^{(3-2n)\ell}-1}{3-2n}.
\end{equation}
Taking $\ell \rightarrow 0$, and considering the dependence of $\nu$ and $D$ with $\Lambda$, \cite{Singha14}, the following differential equation is obtained,
\begin{equation}
    \frac{dl_{n,1}}{d\ell}=\frac{2^{n+1} \Gamma(n-\frac{1}{2})}{4 \pi^{3/2} (n-1)!} \frac{(D \nu \frac{D \lambda^2}{2\pi \nu^3})^{(1-n)/4}}{\Lambda^{\frac{5}{2}(n-1)}(\ell)},
\end{equation}
whose solutions for large $\ell$ become
\begin{equation}
    l_{n,1}(\ell) \simeq \frac{2^{n+1} \Gamma(n-\frac{1}{2})}{4 \pi^{3/2} (n-1)!} \frac{(D \nu \frac{D \lambda^2}{2\pi \nu^3})^{(1-n)/4}}{\frac{5}{2}(n-1) \Lambda^{\frac{5}{2}(n-1)}(\ell)}.
\end{equation}
Due to symmetry among $k_1,\ldots,k_{n-1}$, we take \cite{Yakhot86,Singha14,Singha15,Singha16b}
\begin{equation}
    l_{n,1}(k)=\frac{2^{n+1} \Gamma(n-\frac{1}{2})}{4 \pi^{3/2} (n-1)!} \frac{(D \nu \frac{D \lambda^2}{2\pi \nu^3})^{(1-n)/4}}{\frac{5}{2}(n-1)} \prod_{j=1}^{n-1} \frac{1}{k_j^{5/2}}.
\end{equation}
For $n>1$, as $ k^{5/2} f(\omega/k^z) = k^{-3/2} \nu(k)^{-2} |G(k,\omega)|^{-2}$, where $f$ is a scaling function [$f(u) \rightarrow 1$ as $u \rightarrow 0$], we substitute
$k_i^{-5/2} \simeq k_i^{3/2}  \nu^{2}(k_i) |G(k_i,\omega_i)|^{2}$. Finally,
\begin{equation}
\langle u^n \rangle_c = A \int_{\mathbb{R}^{2(n-1)}} G(k_n,\omega_n) k_n \prod_{i=1}^{n-1} \frac{dk_i d\omega_i}{(2\pi)^2} k_i G(k_i,\omega_i) k_i^{3/2}  \nu^{2}(k_i) |G(k_i,\omega_i)|^{2},
\label{CumulanteFinal}
\end{equation}
where $A=\pi^{n-1/2} {\rm i}^n \Gamma (n-1/2) K 2D / [n!(n-1) \lambda^{n-2}]$.

\begin{figure}[!t]
\begin{center}
\includegraphics[width=0.8\columnwidth]{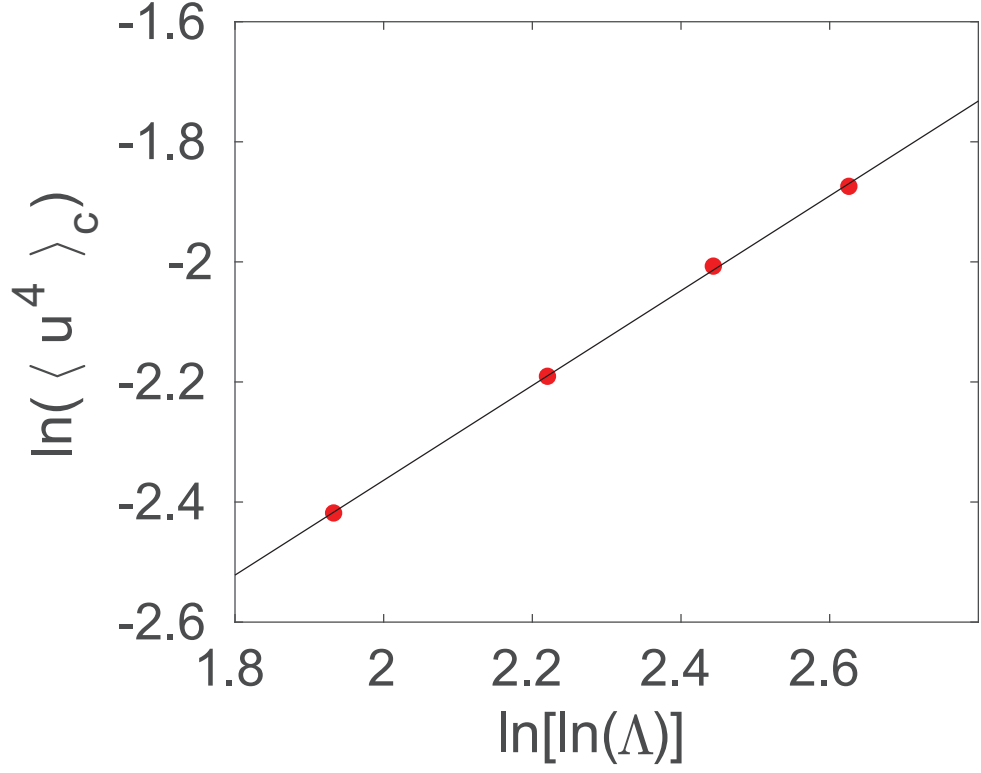} 
\caption{Numerical computation of the fourth cumulant in the $[k_1,k_2,k_3] \in [1,\Lambda]^3$ region, for different values of $\Lambda$ (symbols). The solid line shows a linear fit of the numerical data, and corresponds to the straight line $y=0.79x-3.94$, hence $\langle u^4 \rangle_c \sim (\ln \Lambda)^{0.79}$.}
\label{FigW4}
\end{center}
\end{figure}

\subsubsection{Kurtosis behavior with lattice spacing}

The fourth cumulant of the fluctuation distribution has been estimated for different values of the lattice spacing $s$ by means of analytical integration in $\omega_1,\omega_2,\omega_3$, and numerical integration in $k_1,k_2,k_3$. Parameters have been chosen so as to make $A=1$ and $D\lambda^2/2 \pi=1$. Integration limits in $k_1,k_2,k_3$ of the form $[1,\Lambda]$ have been taken for different values of $\Lambda \propto 1/s$, in order to characterize the divergence of the integral with the lattice spacing $s$. The conclusion is that $\langle u^4 \rangle_c \sim (\ln \Lambda)^{0.79}$, see Fig.\ \ref{FigW4}, a result which is employed in the MT.

\clearpage
\subsection{Further considerations}

\subsubsection{2D generalization of the Burgers equation with conserved noise}

It is natural to consider if non-KPZ behavior also occurs for the Burgers equation with conserved noise in higher dimensions. Note that, if the equation is to be for a scalar field, it can no longer be the derivative of the KPZ equation, as in higher dimensions this is a vector field. Nevertheless, e.g.\ for 2D interfaces a scalar generalization of Eq.\ \eqref{Burgers} can still be formulated in close analogy with the case of non-conserved noise \cite{Rodriguez19}; it reads
\begin{equation}
\partial_t u = \nu \ \nabla^2 u + \lambda u (\partial_x u + \partial_y u) + \partial_x \eta_x + \partial_y \eta_y,
\label{eq:gHK}
\end{equation}
$$
\langle \eta_i(\mathbf{r}_1,t_1) \eta_j(\mathbf{r}_2,t_2) \rangle = 2D \delta_{ij}\delta(\mathbf{r}_1-\mathbf{r}_2) \delta(t_1-t_2), \qquad i,j=x,y.
$$
This continuum model is a conserved-noise version of the equation (the so-called generalized 
Hwa-Kardar (gHK) equation \cite{Vivo14}) shown in \cite{Rodriguez19} to provide the 2D generalization of the 1D Burgers equation with non-conserved noise. In turn, the gHK equation has as a particular case, e.g., the well-known model proposed by Hwa and Kardar to describe the height fluctuations of a running sand pile \cite{Hwa92}.

We next study numerically the scaling exponents and field statistics of Eq.\ \eqref{eq:gHK}. Specifically, in our simulations we employ the same numerical scheme used in the MT. The time evolution of the structure factor $S(k,t)=S(k_x,0,t)=S(0,k_y,t)$ \cite{Rodriguez19,Vivo14} is shown in Fig.\ \ref{fig:Sk2D}. For increasing time and as in 1D, $S(k,t)$ converges towards $k$-independent (i.e., white noise) behavior except for the largest values of $k$, due to the limited accuracy of the numerical scheme at small scales. 
Note the small range of values that occur for $S(k,t)$, leading to large relative numerical errors, which are specially large at such small scales. Data collapse is achieved for $\alpha=-1/2$ and $z=3/2$, notably the same numerical values as in 1D, i.e., for Eq.\ (3) of the MT.
\begin{figure}[!b]
\begin{center}
\includegraphics[width=0.7\columnwidth]{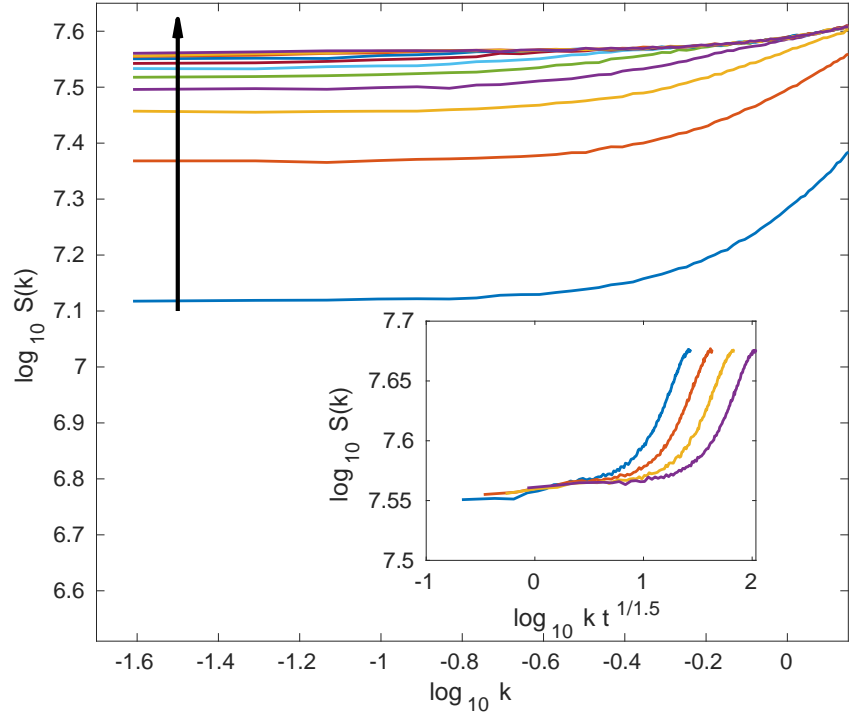} 
\caption{Time evolution of the structure factor $S(k)=S(k_x,0)=S(0,k_y)$ for the gHK equation with conserved noise, Eq.\ \eqref{eq:gHK}. Time increases in the direction of the arrow, doubling for each consecutive live starting at $t_0=1$. The inset show the collapse of the $k^{2\alpha+1}S(k t^{1/z})$ curves for the longest times within the nonlinear regime, using exponent values $\alpha=-1/2$ and $z=3/2$. Here, $L_x=L_y=256$, $\Delta x=\Delta y =1$, $\Delta t=0.01$, and $\nu=D=\lambda=1$.}
\label{fig:Sk2D}
\end{center}
\end{figure}

The one-point statistics of the $u$ field has been also numerically characterized for Eq.\ \eqref{eq:gHK}. The time evolution of the skewness and excess kurtosis, as well as the histograms for $X$ as defined in Eq.\ (4) of the MT, are plotted in Fig.\ \ref{fig:Hist2D}, both in the growth regime and after saturation to steady state. The variations in $u$ show a symmetric PDF and behave not far from a Gaussian random variable. The large impact of numerical errors for these data related with the limited range of values of the structure factor (see also Fig.\ \ref{fig:Sk2D}) seems to induce a slight negative excess kurtosis.

\begin{figure}
\begin{center}
\includegraphics[width=0.9\columnwidth]{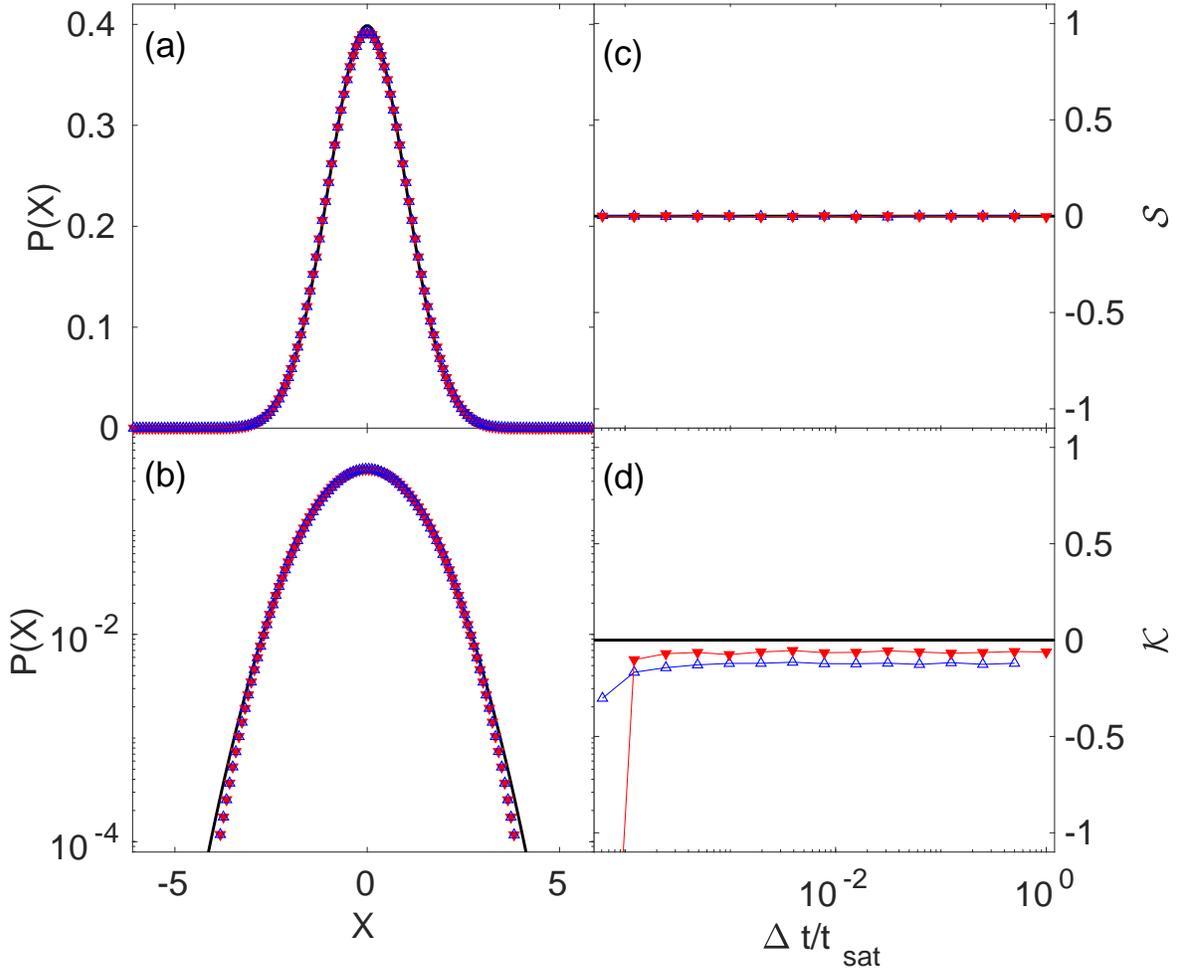}
\caption{Histogram for $X(x,\Delta t,t_0)$ (with $\phi=u$) (a,b) from simulations of Eq.\ \eqref{eq:gHK} with $\nu=D=\lambda=1$, and $L_x=L_y=1024$. In all panels blue up (red down) and empty (filled) triangles correspond to the growth (saturation) regime, with black solid lines showing exact Gaussian values; $t_{sat}=64$, and $\Delta t=0.1$ are used. All units are arbitrary.}
\label{fig:Hist2D}
\end{center}
\end{figure}

\subsubsection{Exact linear approximation}

Finally, let us remark that the Gaussian nature of the field PDF displayed by Eq.\ \eqref{Burgers} in its large-scale nonlinear regime allows for an exact Gaussian (asymptotic) approximation of the equation in terms of a {\em linear} model which is in the same universality class, including scaling exponent values {\em and} Gaussian statistics. Again, this is akin to the case of the scalar Burgers equation with non-conserved noise \cite{Rodriguez19}, also including higher-dimensional and strongly anisotropic generalizations \cite{Vivo12,Vivo14}, like the celebrated Hwa-Kardar equation for the height of a running sand pile \cite{Hwa92}. In contrast with these cases, Eq.\ \eqref{Burgers} does {\em not} support hyperscaling ($2\alpha+d=z$) \cite{Krug97,Barabasi95}, hence noise correlations are required in order to match the full universal behavior. Specifically, the linear, non-local equation
\begin{eqnarray}
 & \partial_t \tilde{u}(k,t) = -|k|^{3/2} \tilde{u}(k,t) + \tilde{\eta}(k,t), & \\
\label{Proxy}
 & \langle \tilde{\eta}(k,t) \tilde{\eta} (k',t') \rangle = |k|^{3/2} \delta (k+k') \delta (t-t'), &
\end{eqnarray}
yields the exact same asymptotic behavior of the nonlinear Eq.\ \eqref{Burgers}. Note that a similar exact Gaussian approximation is not possible for systems with non-Gaussian statistics (like the KPZ equation), not even considering correlations in the noise.

\vspace{0.5 cm}

\end{widetext}


\footnotesize

\begin{thebibliography}{99}

\bibitem{Kardar86}
M. Kardar, G. Parisi, and Y.-C. Zhang,
Dynamic Scaling of Growing Interfaces,
Phys. Rev. Lett. {\bf 56}, 889 (1986).

\bibitem{Taeuber14}
U. C. T\"{a}uber, {\em Critical Dynamics} (Cambridge University Press, Cambridge, England, 2014).

\bibitem{Livi17}
R. Livi and P. Politi, {\em Nonequilibrium Statistical Physics: A Modern Perspective} (Cambridge University Press, Cambridge, England, 2017).

\bibitem{Lenz20}
W. Lenz,
Beitr\"age zum Verst\"andnis der magnetischen Eigenschaften in festen K\"orpern",
Phys. Z. {\bf 21}, 613 (1920).

\bibitem{Onsager44}
L. Onsager,
Crystal Statistics. I. A Two-Dimensional Model with an Order-Disorder Transition,
Phys. Rev. {\bf 65}, 117 (1944).

\bibitem{Sasamoto10}
T. Sasamoto and H. Spohn,
One-Dimensional Kardar-Parisi-Zhang Equation: An Exact Solution and its Universality,
Phys. Rev. Lett. {\bf 104}, 230602 (2010).

\bibitem{Amir10}
G. Amir, I. Corwin, and J. Quastel,
Probability Distribution of the Free Energy of the Continuum Directed Random Polymer in 1+1 Dimensions,
Commun. Pure Appl. Math. {\bf 64}, 466 (2011).

\bibitem{Calabrese11}
P. Calabrese and P. Le Doussal,
Exact Solution for the Kardar-Parisi-Zhang Equation with Flat Initial Conditions,
Phys. Rev. Lett. {\bf 106}, 250603 (2011).

\bibitem{Henkel99}
M. Henkel, {\em Conformal Invariance and Critical Phenomena} (Springer-Verlag, Berlin, 1999).

\bibitem{Henkel08}
M. Henkel, H. Hinrichsen, and S. L\"ubeck, {\em Non-Equilibrium Phase Transitions, Vol. I}
(Science + Business Media, Dordrecht, 2008).

\bibitem{Halpin-Healy15}
T. Halpin-Healy and K. A. Takeuchi,
A KPZ Cocktail-Shaken, not Stirred...,
J. Stat. Phys. \textbf{160}, 794 (2015).

\bibitem{Takeuchi18}
K. A. Takeuchi,
An appetizer to modern developments on the Kardar-Parisi-Zhang universality class,
Physica A {\bf 504}, 77 (2018).

\bibitem{Hallatschek07}
O. Hallatschek {\em et al.}, %Hersen, Pascal, Ramanathan, Sharad, Nelson, David R.,
Genetic drift at expanding frontiers promotes gene segregation,
Proc. Natl. Acad. Sci. USA {\bf 104}, 19926 (2007).

\bibitem{Takeuchi11}
K. A. Takeuchi {\em et al.}, %M. Sano, T. Sasamoto, and H. Spohn,
Growing interfaces uncover universal fluctuations behind scale invariance,
Sci. Rep. {\bf 1}, 34 (2011).

\bibitem{Beijeren12}
H. van Beijeren,
Exact results for anomalous transport in one-dimensional hamiltonian systems,
Phys. Rev. Lett. {\bf 108}, 180601 (2012).

\bibitem{Mendl13}
C. B. Mendl and H. Spohn,
Dynamic correlators of fermi-pasta-ulam chains and nonlinear fluctuating hydrodynamics,
Phys. Rev. Lett. {\bf 111}, 230601 (2013).

\bibitem{Yunker13}
P. J. Yunker {\em et al.}, %Lohr, Matthew A., Still, Tim, Borodin, Alexei, Durian, D. J., Yodh, A. G.
Effects of Particle Shape on Growth Dynamics at Edges of Evaporating Drops of Colloidal Suspensions,
Phys. Rev. Lett. {\bf 110}, 035501 (2013).

\bibitem{Almeida14}
R. A. L. Almeida, {\em et al.} %Ferreira, S. O., Oliveira, T. J., & Reis, F. D. A. A.
Universal fluctuations in the growth of semiconductor thin films,
Phys. Rev. B {\bf 89}, 045309 (2014).

\bibitem{Orrillo17}
P. A. Orrillo, {\em et al.} %Santalla, S. N., Cuerno, R., Vázquez, L., Ribotta, S. B., Gassa, L. M., Vela, M. E.
Morphological stabilization and KPZ scaling by electrochemically induced co-deposition of nanostructured NiW alloy films,
Sci. Rep. {\bf 7}, 17997 (2017).

\bibitem{Nesic14}
S. Nesic, R. Cuerno, and E. Moro,
Macroscopic response to microscopic intrinsic noise in three-dimensional Fisher fronts,
Phys. Rev. Lett. {\bf 113}, 180602 (2014).

\bibitem{Santalla15}
S. N. Santalla {\em et al.}, %J. Rodríguez-Laguna, T. Lagatta, and R. Cuerno, R.
Random geometry and the Kardar-Parisi-Zhang universality class,
New J. Phys. {\bf 17}, 33018 (2015).

\bibitem{Altman15}
E. Altman, {\em et al.}, %Sieberer, L. M., Chen, L., Diehl, S., & Toner, J.
Two-dimensional superfluidity of exciton polaritons requires strong anisotropy,
Phys. Rev. X {\bf 5}, 011017 (2015).

\bibitem{Chen16}
L. Chen {\em et al.}, %Lee, C. F., & Toner, J..
Mapping two-dimensional polar active fluids to two-dimensional soap and one-dimensional sandblasting,
Nature Comm. {\bf 7}, 12215 (2016).

\bibitem{Nahum17}
A. Nahum {\em et al.}, %Ruhman, J., Vijay, S., & Haah, J.
Quantum entanglement growth under random unitary dynamics,
Phys. Rev. X {\bf 7}, 031016 (2017).

\bibitem{Kriechebauer10}
T. Kriecherbauer and J. Krug,
A pedestrian's view on interacting particle systems, KPZ universality and random matrices,
J. Phys. A: Math. Theor. {\bf 43}, 403001 (2010).

\bibitem{Fortin15}
J.-Y. Fortin, and M. Clusel,
Applications of Extreme Value Statistics,
J. Phys. A: Math. Theor. {\bf 48}, 183001 (2015).

\bibitem{Forster77} D. Forster, D. R. Nelson, and M. J. Stephen,
Large-distance and long-time properties of a randomly stirred fluid,
Phys. Rev. A \textbf{16}, 732 (1977).

\bibitem{DaPrato94}
G. Da Prato, A. Debussche, and R. Temam,
Stochastic Burgers' equation,
Nonlin. Diff. Eq. Appl. {\bf 1}, 389 (1994).

\bibitem{Bertini97}
L. Bertini and G. Giacomin,
Stochastic Burgers and KPZ Equations from Particle Systems,
Comm. Math. Phys. {\bf 183}, 571 (1997).

\bibitem{Gubinelli13}
M. Gubinelli and M. Jara,
Regularization by noise and stochastic Burgers equations,
Stoch. PDE Anal. Comp. {\bf 1}, 325 (2013).

\bibitem{Frisch95}
U. Frisch, {\em Turbulence: the legacy of A. N. Kolmogorov} (Cambridge University Press, Cambridge, England, 1995).

\bibitem{Krommes02}
J. A. Krommes,
Fundamental statistical descriptions of plasma turbulence in magnetic fields,
Phys. Rep. {\bf 360}, 1 (2002).

\bibitem{Spohn91}
H. Spohn, {\em Large Scale Dynamics of Interacting Particles} (Springer-Verlag, Berlin, 1991).

\bibitem{Krug97}
J. Krug,
Origins of scale invariance in growth processes,
Adv. Phys. \textbf{46}, 139 (1997).

\bibitem{Clusel08}
M. Clusel and E. Bertin,
Global fluctuations in physical systems a subtle interplay between sum and extreme value statistics,
Int. J. Mod. Phys. B {\bf 22}, 3311 (2008).

\bibitem{Ueno05}
K. Ueno, H. Sakaguchi, and M. Okamura,
Renormalization-group and numerical analysis of a noisy Kuramoto-Sivashinsky equation in 1+1 dimensions,
Phys. Rev. E \textbf{71}, 046138 (2005).

\bibitem{Hairer11}
M. Hairer and J. Voss,
Approximations to the Stochastic Burgers Equation,
J. Nonlinear. Sci. \textbf{21}, 897 (2011).

\bibitem{Sasamoto09}
T. Sasamoto and H. Spohn,
Superdiffusivity of the 1D lattice Kardar-Parisi-Zhang equation,
J. Stat. Phys. {\bf 137}, 917 (2009).

\bibitem{Barabasi95}
A.-L. Barab\'asi and H. E. Stanley, \emph{Fractal concepts in surface growth} (Cambridge University Press, Cambridge, England, 1995).

\bibitem{Takeuchi13}
K. A. Takeuchi,
Crossover from Growing to Stationary Interfaces in the Kardar-Parisi-Zhang Class,
Phys. Rev. Lett. \textbf{110}, 210604 (2013).

\bibitem{Squizzato18}
D. Squizzato, L. Canet, and A. Minguzzi,
Kardar-Parisi-Zhang universality in the phase distributions of one-dimensional exciton-polaritons,
Phys. Rev. B \textbf{97}, 195453 (2018).

\bibitem{Bornemann10}
F. Bornemann,
On the numerical evaluation of Fredholm determinants,
Math. Comput. {\bf 79} 871 (2010).

\bibitem{Prahofer04}
M. Pr\"ahofer and H. Spohn,
Exact Scaling Functions for One-Dimensional Stationary KPZ Growth,
J. Stat. Phys. \textbf{115}, 255 (2004).

\bibitem{Imamura12}
T. Imamura and T. Sasamoto,
Exact Solution for the Stationary Kardar-Parisi-Zhang Equation,
Phys. Rev. Lett. {\bf 108}, 190603 (2012).

\bibitem{Singha14} T. Singha and M. K. Nandy,
Skewness in (1 + 1)-dimensional Kardar-Parisi-Zhang-type growth,
Phys. Rev. E \textbf{90}, 062402 (2014).

\bibitem{Singha15} T. Singha and M. K. Nandy,
Kurtosis of height fluctuations in (1 + 1) dimensional KPZ Dynamics,
J. Stat. Mech.: Theor. Exp. (2015), 05020.

\bibitem{Singha16} T. Singha and M. K. Nandy,
Hyperskewness of (1+1)-dimensional KPZ height fluctuations,
J. Stat. Mech.: Theor. Exp. (2016), 013203.

\bibitem{Singha16b}
T. Singha and M. K. Nandy,
A renormalization scheme and skewness of height fluctuations in (1 + 1)-dimensional VLDS dynamics,
J. Stat. Mech.: Theor. Exp. 023205 (2016).

\bibitem{Rodriguez19}
E. Rodriguez-Fernandez and R. Cuerno,
Gaussian statistics as an emergent symmetry of the stochastic scalar Burgers equation,
Phys. Rev. E \textbf{99}, 042108 (2019).

\bibitem{Yakhot86}
V. Yakhot and S. A. Orszag,
Renormalization-Group Analysis of Turbulence,
Phys. Rev. Lett. \textbf{57}, 1722 (1986).

\bibitem{Smith17}
N. R. Smith, B. Meerson, and P. V. Sasorov,
Local average height distribution of fluctuating interfaces,
Phys. Rev. E {\bf 95}, 012134.  (2017).

\bibitem{Carrasco16}
I. S. S. Carrasco and T. J. Oliveira,
Universality and geometry dependence in the class of the nonlinear molecular beam epitaxy equation,
Phys. Rev. E {\bf 94}, 050801(R) (2016).

\bibitem{Burgers74}
J. M. Burgers, {\em The Nonlinear Diffusion Equation} (D. Reidel,
Dordrecht, Holland, 1974).

\bibitem{Bendaas18}
S. Bendaas,
Periodic wave shock solutions of Burgers equations,
Cogent Math. Stat. {\bf 5}, 1463597 (2018).

\bibitem{Sakaguchi03}
For a study of Eq.\ \eqref{eq:burgers} with step boundary conditions, see
H. Sakaguchi,
Shock Structures and Velocity Fluctuations in the Noisy Burgers and KdV-Burgers Equations,
Prog. Theor. Phys. {\bf 110}, 187 (2003).

\bibitem{Vivo12}
E. Vivo, M. Nicoli, and R. Cuerno,
Strong anisotropy in two-dimensional surfaces with generic scale invariance: Gaussian and related models,
Phys. Rev. E {\bf 86}, 051611 (2012).

\bibitem{Vivo14}
E. Vivo, M. Nicoli, and R. Cuerno,
Strong anisotropy in two-dimensional surfaces with generic scale invariance: Nonlinear effects,
Phys. Rev. E \textbf{89}, 042407 (2014).

\bibitem{Hwa92} T. Hwa and M. Kardar,
Avalanches, hydrodynamics, and discharge events in models of sandpiles,
Phys. Rev. A \textbf{45}, 7002 (1992).

\bibitem{Stoyanov97}
J. Stoyanov,
{\em Counterexamples
in Probability}
(John Wiley and Sons, Chichester, 1997).






\end{thebibliography}
\end{document}